\documentclass[
  aps,
  prb,
  twocolumn,
  superscriptaddress
]{revtex4-2}
\usepackage{amsmath,amssymb,bm}
\usepackage{graphicx}
\usepackage{dcolumn}
\usepackage{color}
\usepackage{accents}
\usepackage{mathtools}
\usepackage[normalem]{ulem}
\usepackage{url}
\usepackage[ 
  colorlinks=true,    
  linkcolor=blue,     
  citecolor=blue,     
  urlcolor=blue,      
]{hyperref}

\begin{document}

\title{Low-temperature anomaly and anisotropy of critical magnetic fields in transition-metal dichalcogenide superconductors}
\date{\today}

\author{Tomoya Sano}
\affiliation{Department of Applied Physics, Hokkaido University, Sapporo 060-8628, Japan}
\author{Kota Tabata}
\affiliation{Department of Applied Physics, Hokkaido University, Sapporo 060-8628, Japan}
\author{Akihiro Sasaki}
\affiliation{Department of Applied Physics, Hokkaido University, Sapporo 060-8628, Japan}
\author{Yasuhiro Asano}
\affiliation{Department of Applied Physics, Hokkaido University, Sapporo 060-8628, Japan}

\begin{abstract}
We clarify why spin-singlet superconductivity persists 
in monolayer transition-metal dichalcogenides even 
in high magnetic fields beyond the Pauli limit.
The phenomenon called Ising protection is caused by
two magnetically active potentials: a Zeeman field and an Ising spin-orbit interaction.
These potentials induce two spin-triplet pairing correlations in a spin-singlet superconductor. 
One belonging to odd-frequency symmetry class arises solely from a Zeeman field and  
always makes the superconducting state unstable.
The other belonging to even-frequency symmetry class arises from the interaction between 
the two magnetic potentials and eliminates the instability caused by odd-frequency pairs.
The presence or absence of such even-frequency spin-triplet pairs
explains the anisotropy of the Ising protection.
The analytical expression of the superfluid weight 
enables us to conclude that induced even-frequency spin-triplet Cooper pairs
support spin-singlet superconductivity in high Zeeman fields.
\end{abstract}

\maketitle

\section{Introduction} \label{sec:introduction}
Superconductivity is quenched by an external magnetic field in two ways: 
orbital effect fluctuates the phase of superconducting condensate 
and spin Zeeman effect breaks a spin-singlet Cooper pair.
In monolayer transition metal dichalcogenide (TMD) superconductors 
such as MoS$_{2}$, NbSe$_{2}$, TaS$_{2}$~\cite{saito:natphys2016,xi:natphys2016,barrera:natcom2018,simon:natcom2024}, 
the orbital effects are negligible for a magnetic field applied parallel to the monolayer.
The critical magnetic field in such a case is limited 
only by the Zeeman effect and its value at zero temperature is called 
Pauli limit $H_{p}$~\cite{chandrasekhar:apl1962,clogston:prl1962}.
In experiments, however, observed critical fields tend to go over the Pauli limit 
in various superconductors (SCs) such as a thin Al film \cite{tedrow:prb1982}, 
organic SCs~\cite{agosta:prb2012,lee:prb2000}, 
noncentrosymmetric SCs~\cite{bauer:prl2004,kimura:prl2007,hoshi:sr2022}, 
and TMDs~\cite{foner:pra1973,prober:prb1980}.
Theories have shown that the spin-orbit interactions (SOIs) weaken 
the pair-breaking effect by a Zeeman field~\cite{maki:ptp1964,klemm:prb1975,tedrow:prb1982}. 
In TMDs, the Ising-type spin-orbit interaction (SOI) locks spin of an electron 
along the perpendicular direction to the monolayer~\cite{lu:science2015}.
A spin-singlet pair consisting of such two electrons seems to be robust 
under Zeeman field parallel to the monolayer. 
This phenomenon is called Ising protection of superconductivity in recent contexts.
The experiments on various SCs~\cite{foner:pra1973,prober:prb1980,lee:prb2000,bauer:prl2004,kimura:prl2007,agosta:prb2012,lu:science2015,hoshi:sr2022} 
have also suggested that the Ising protection tends to be stronger at lower temperatures.
Although existing theories reproduce such tendency 
in $H$-$T$ phase diagrams~\cite{frigeri:prl2004,saito:natphys2016,uher:prb1986}, 
there is no satisfactory explanation for the low-temperature anomaly in the Ising protection.
In addition, the remarkable anisotropy of the Ising protection has not yet been well understood. 
The purpose of this paper is to provide a physical picture that 
fully explains these characteristics of the phenomenon.
For the purpose, 
we pay special attention to the frequency-symmetry of Cooper pair 
that are induced by the magnetic potentials.

Odd-frequency superconductivity, pair potential belongs to odd-frequency symmetry class,
has been studied~\cite{kirkpatrick:prl1991,balatsky:prb1992}
since the concept of odd-frequency Cooper pairing was introduced
to explain superfluidity of $^{3}$He~\cite{berezinskii:jetplett1974}.
Since odd-frequency Cooper pairs indicate the paramagnetic response to magnetic fields,
the uniform odd-frequency superconductivity is unstable thermodynamically~\cite{fominov:prb2015}.
However,  odd-frequency Cooper pairs can exist as an induced subdominant pairing correlation
in a SC  in which the pair potential belongs to usual even-frequency symmetry class.
Their presence has been well established at superconducting proximity structures,
vortex cores, surfaces of unconventional SCs, and various impurities~\cite{bergeret:prl2001,bergeret:rmp2005,tanaka:prl2007,tanaka:jpsj2012,linder:rmp2019,cayao:epjst2020}.
In these cases, odd-frequency Cooper pairs cause unusual phenomena
such as the anomalous surface impedance of superconducting proximity structures~\cite{asano:prl2011}
and the paramagnetic response of small unconventional SCs~\cite{suzuki:prb2014}.
A theory~\cite{schaffer:prb2013} has suggested that odd-frequency Cooper pairs exist 
even in a uniform superconducting condensate when an electron has extra degree of freedom 
such as bands, orbitals, spins, and sublattices.
As odd-frequency Cooper pairs decrease the superfluid density,
they decrease the transition temperature to a superconducting phase~\cite{asano:prb2015,ramires:prb2018,triola:annalen2020}.

In monolayer TMDs, an electron has an extra internal degree of freedom called as valleys, 
which enables an extra symmetry option in a pair correlation function 
in addition to the conventional three symmetry options: 
frequency, spin-structure, and momentum-parity. 
The parity of a pairing function under the interchange of the valley 
indices also characterizes the symmetry of a Cooper pair. 
We analytically solve the Gor'kov equation for the Bogoliubov-de Gennes Hamiltonian which 
describes the superconducting states 
in the presence of such an extra internal degree of freedom of an electron.
The symmetry of Cooper pairs is discussed by 
using the analytical expression of the anomalous Green's function. 
The transition temperature is obtained by solving the gap equation numerically. 
In spin-singlet SCs, a Zeeman field always makes the superconducting state unstable 
because it generates odd-frequency spin-triplet Cooper pairs and decreases the superfluid density.
Ising SOI generates an extra pairing correlation 
that belongs to even-frequency spin-triplet odd-valley symmetry class. 
We will conclude that such induced Cooper pairs remove 
the instability of superconductivity due to a Zeeman field alone.
The anisotropy of the Ising protection is characterized 
by the presence or absence of the extra pairing correlation.

This paper is organized as follows.
We explain our theoretical model and 
provide the analytical expression of the anomalous Green's function in Sec.~\ref{sec:model}.
The gap equation and $H$-$T$ phase diagram are presented in Sec.~\ref{sec:gap_equation}.
We discuss the superfluid density in Sec.~\ref{sec:superfluid_density}. 
In Sec.~\ref{sec:discussion}, we discuss the influences of Rashba SOI and 
impurities involving SOI on the critical fields.
The conclusion is given in Sec.~\ref{sec:conclusion}.
We use the unit of $k_{B} = \hbar = c = 1$, where $k_{B}$ is the Boltzmann constant 
and $c$ is the speed of light.

\begin{table*}[ttt]
\caption{
Symmetry classification of the pair correlation functions.
The first column specifies the origin of each pairing correlation.
The top row represents the principal pairing correlation 
that is linked to the pair potential through Eq.~\eqref{eq:gap_equation}.
The remaining pairing correlations are induced by magnetically active potentials.
The SOI generates the pairing correlation at the second row from the top.
A Zeeman field induces the two pairing correlations: 
one belongs to odd-frequency symmetry as shown in the third row and 
the other belongs to even-frequency symmetry as listed at the bottom row.
The last column denotes the superfluid weight 
in Eqs.~\eqref{eq:q_comp1} and \eqref{eq:q_comp2} to which the pairing correlation contributes.
}
\begin{ruledtabular}
\begin{tabular}{cccccc}
 pairing origin & frequency & spin ($\times i \hat{\sigma}_y$) & parity & valley-parity & component of superfluid weight\\
\colrule
 principal & even & singlet & even & even $\hat{\rho}_x$ & $q_0$\\
 induced & even & triplet $\bm{\beta} \cdot \hat{\bm{\sigma}}$ & even  & odd $\hat{\rho}_y$ & $q_0$ \\
 induced & odd & triplet $\bm{H} \cdot \hat{\bm{\sigma}}$ & even & even $\hat{\rho}_x$ & $q_{\mathrm{odd}}$\\
 induced & even & triplet $\bm{\beta} \times \bm{H}  \cdot \hat{\bm{\sigma}}$ & even  & odd $\hat{\rho}_y$ & $q_{\perp}$\\
\end{tabular}
\end{ruledtabular}
\label{table1}
\end{table*}

\section{Model} \label{sec:model}
In typical TMDs, the Fermi surfaces surround $K$ and $K^{\prime}$ points in the Brillouin zone~\cite{saito:natphys2016,xi:natphys2016,barrera:natcom2018}. 
We describe such two Fermi surfaces in terms of two valleys. 
The normal state Hamiltonian is given by
\begin{align}
  \check{H}_{\mathrm{N}} &(\bm{k}) \label{eq:normal state hamiltonian}
  =
  \hat{\xi}_{\bm{k}} \hat{\sigma}_{0}
  +
  \bm{\beta} \cdot \hat{\bm{\sigma}} \hat{\rho}_{z}
  +
  \mu_{\mathrm{B}} \bm{H} \cdot \hat{\bm{\sigma}} \hat{\rho}_{0} , \\
  \hat{\xi}_{\bm{k}} \label{eq:xipm}
  &=
  \frac{1}{2 m} \left( \bm{k} \hat{\rho}_{0} - \bm{K} \hat{\rho}_{z} \right)^{2} - \mu ,  
\end{align}
where $\mu$ is the chemical potential, $\bm{K}$ is a vector connecting $K$ and $\Gamma$ points, and $\bm{H}$ represents a Zeeman field with $\mu_{\mathrm{B}}$ being the Bohr magneton.
The Pauli matrices in spin and valley spaces are denoted by $\hat{\bm{\sigma}} = (\hat{\sigma}_{x}, \hat{\sigma}_{y}, \hat{\sigma}_{z})$ and $\hat{\bm{\rho}} = (\hat{\rho}_{x}, \hat{\rho}_{y}, \hat{\rho}_{z})$, respectively.
The unit matrices in the two spaces are denoted by $\hat{\sigma}_{0}$ and $\hat{\rho}_{0}$.
The notation $\check{\cdots}$ denotes a $4\times4$ matrix
in the combined spin and valley spaces.
We assume that the SOI $\bm{\beta}$ is independent of $\bm{k}$ and changes its sign in the two valleys.
The time-reversal operation in such model is given by $\mathcal{T} = i \hat{\sigma}_{y} \hat{\rho}_{x} \mathcal{K}$, where $\mathcal{K}$ means the complex conjugation plus $\bm{k} \to - \bm{k}$.
The particle-hole conjugation is represented by $\undertilde{\check{H}}_{\mathrm{N}} (\bm{k}) = \check{H}^{*}_{\mathrm{N}} (- \bm{k})$ as usual.
Two electrons at the different valleys form a spin-singlet $s$-wave Cooper pair.
The pair potential for such a pair is represented as
\begin{align}
  \check{\Delta} = \Delta \, i \, \hat{\sigma}_{y} \, \hat{\rho}_{x}.
\end{align}
We solve the Gor'kov equation for the Bogoliubov-de Gennes Hamiltonian
\begin{align}
  &\left[ i \omega_{n} - H_{\mathrm{BdG}} (\bm{k}) \right]
  \begin{bmatrix}
    \check{\mathcal{G}} & 
    \check{\mathcal{F}} \\
    - \undertilde{\check{\mathcal{F}}} & 
    - \undertilde{\check{\mathcal{G}}} 
  \end{bmatrix}_{(\bm{k}, \omega_{n})}
  = 1 ,\label{eq:bdg} \\ 
  &H_{\mathrm{BdG}} (\bm{k})
  =
  \begin{bmatrix}
    \check{H}_{\mathrm{N}} (\bm{k}) & \check{\Delta} \\
    - \undertilde{\check{\Delta}} & - \undertilde{\check{H}}_{\mathrm{N}} (\bm{k})
  \end{bmatrix} ,
\end{align}
where $\omega_{n} = (2n + 1) \pi T$ is a Matsubara frequency with $T$ being a temperature.
The Hamiltonian is block-diagonalized into two particle-hole spaces. 
An electron around $K$ and a hole around $K^{\prime}$ are coupled in one particle-hole space, while an electron around $K^{\prime}$ and a hole around $K$ are coupled in the other.
The exact solution of the Gor'kov equation is given in Appendix~\ref{sec:solution}. 
$ \bm{K} \hat{\rho}_{z}$ represents the shift of the band bottom from $\Gamma$ point to $K$ and 
 $K^{\prime}$ points. 
As shown in Appendix~\ref{sec:solution}, the shift of the wavenumber by $\bm{K}$ does not affect the stability of the superconducting states because the shift 
is absorbed by changing the range of the summation over $\bm{k}$.  
In what follows, we neglect $\bm{K}$ in the dispersion
and discuss the symmetry of Cooper pairs. 
The anomalous Green's function is given by
\begin{align}
  &\check{\mathcal{F}} (\bm{k}, \omega_{n}) \label{eq:F} \nonumber \\
  &=
  \check{Z}^{-1} \Big[
    - \left( 
      \xi^{2}_{\bm{k}} + \omega^{2}_{n} + \Delta^{2} + \bm{\beta}^{2} - \mu_{\mathrm{B}}^2\bm{H}^{2} 
    \right) \hat{\sigma}_{0} \hat{\rho}_0 \nonumber \\
    &\qquad +
    2 \, \xi_{\bm{k}} \, \bm{\beta} \cdot \hat{\bm{\sigma}} \, \hat{\rho}_{z}
    +2 i \, \omega_{n} \mu_{\mathrm{B}} \bm{H} \cdot \hat{\bm{\sigma}} \hat{\rho}_0 \nonumber \\
    &\qquad +
    2 i \, \left( \bm{\beta} \times \mu_{\mathrm{B}} \bm{H} \right) \cdot \hat{\bm{\sigma}} \hat{\rho}_{z}
  \Big]
  \Delta \left( i \hat{\sigma}_{y} \right) \hat{\rho}_{x} , \\
  &\check{Z} 
  =
  \hat{\sigma}_0 \Big[
  \left(
    \xi^{2}_{\bm{k}} + \omega^{2}_{n} + \Delta^{2} + \bm{\beta}^{2} - \mu_{\mathrm{B}}^{2}\bm{H}^2
  \right)^{2} \hat{\rho}_0 \nonumber \\
  &\quad - 
  4 \left(
    i \omega_{n} \mu_{\mathrm{B}} \bm{H} \hat{\rho}_0 + \xi_{\bm{k}} \bm{\beta} \hat{\rho}_{z}
  \right)^{2} +
  4 \mu_{\mathrm{B}}^2 \left( \bm{\beta} \times \bm{H} \right)^{2} \hat{\rho}_0
  \Big] .\label{eq:z_def}
\end{align}
with a scaler dispersion $\xi_{\bm{k}}= \bm{k}^2/(2m) - \mu$.
There are four symmetry options for a Cooper pair in the present model: frequency, spin configuration, momentum parity, and valley parity. 
In the presence of in-plane magnetic field ($\bm{H} \parallel \bm{x}$ or $\bm{y}$) and the Ising SOI ($\bm{\beta} \parallel \bm{z}$), the two vectors are perpendicular to each other, 
(i.e., $\bm{\beta} \perp \bm{H}$). 
In such a case, 
$\check{Z} $ in Eq.~\eqref{eq:z_def}  is reduced to a unit matrix
\begin{align}
  \check{Z} =
  \hat{\sigma}_0 \hat{\rho}_0 &\Big[
  \left(
    \xi^{2}_{\bm{k}} + \omega^{2}_{n} + \Delta^{2} + \bm{\beta}^{2} - \mu_{\mathrm{B}}^{2}\bm{H}^2
  \right)^{2} \nonumber \\
  &+ 
  4 \omega_{n}^2 \mu_{\mathrm{B}}^2 \bm{H}^2
  - 4 \xi_{\bm{k}}^2 \bm{\beta}^2 
  +
  4 \mu_{\mathrm{B}}^2 \bm{H}^2 \bm{\beta}^2 
  \Big]. 
\end{align}
Symmetry of four pairing correlations in Eq.~\eqref{eq:F} are summarized in Table~\ref{table1}.
The first term in Eq.~\eqref{eq:F} represents 
the pairing correlation that belongs to 
even-frequency spin-singlet $s$-wave even-valley symmetry and 
is linked to the pair potential through the gap equation.
Namely, the attractive interaction between two electrons works in this channel and 
forms the spin-singlet pair potential. 
Such pairing correlation is referred to as \textsl{principal} pairing correlation in this paper.
The SOI induces the pairing correlation at the second term 
which belongs to even-frequency spin-triplet $s$-wave odd-valley symmetry. 
Such even-frequency Cooper pairs always stabilize the superconducting states in a spin-singlet superconductor.
This term explains the anisotropy of the magnetic susceptibility in $s$-wave SCs~\cite{frigeri:njp2004,kim:jpsj2023} and 
the Josephson coupling of a spin-singlet SC to a spin-triplet SC~\cite{yamaki:prb2025}. 
The third term represents the pairing correlation belonging to odd-frequency 
spin-triplet $s$-wave even-valley symmetry. 
Such odd-frequency pairs suppress the critical temperature in Zeeman field because they
reduce the superfluid density and make the superconducting 
state unstable~\cite{maki:ptp1964,sarma:jpcs1963,sato:prb2024}.
The last term represents the pairing correlation belonging to even-frequency spin-triplet $s$-wave 
odd-valley symmetry class. 
The interplay between the SOI and the Zeeman field induces this pairing correlation.
The three types of spin-triplet Cooper pair exist only as subdominant pairing correlations and
do not form any pair potentials. Such pairing correlation is referred to as \textsl{induced} pairing correlation 
as indicated at the first column in Table~\ref{table1}.

In the succeeding sections, we will discuss how these pairing correlations 
affect the critical magnetic field.

\section{Gap equation} \label{sec:gap_equation}
\begin{figure*}[tttt]
  \centering
  \includegraphics[width = 2\columnwidth]{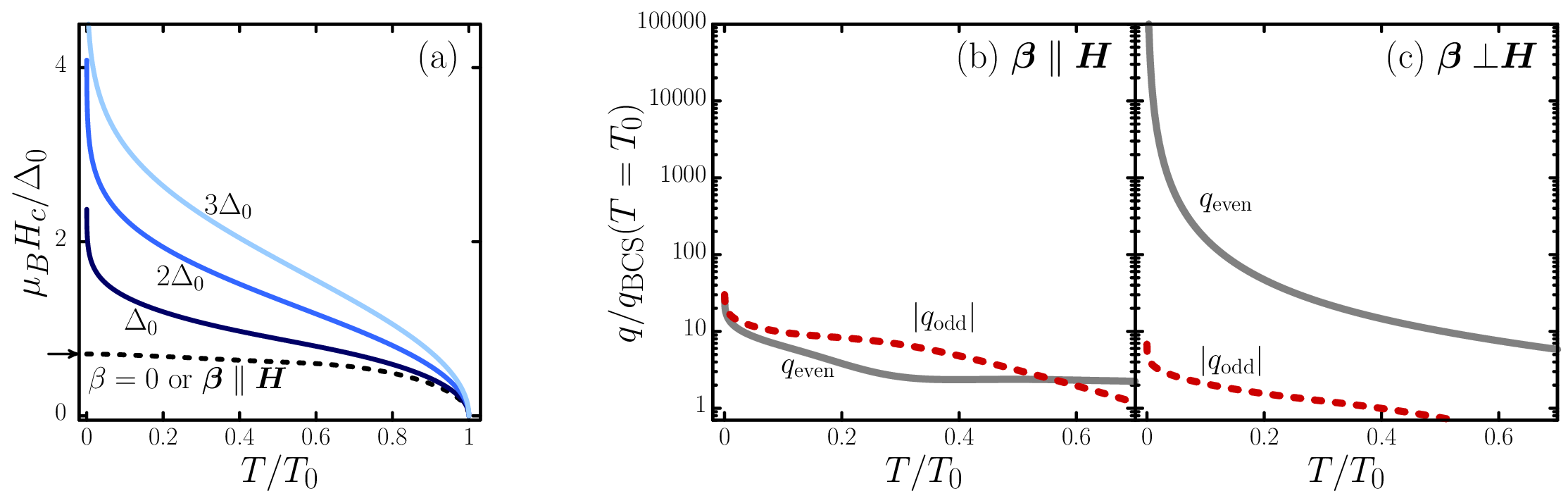}
  \caption{
    The critical magnetic field $H_{c}$ is plotted as a function of temperature $T$ for several values of the Ising spin-orbit interactions $\beta$ in (a).
    The arrow on the vertical axis indicates the Pauli limit $\mu_{\mathrm{B}} H_{p} = \Delta_{0}/\sqrt{2}$.
    The superfluid weights along the transition line in (a) is shown for $\bm{\beta} \parallel \bm{H}$ in (b) 
	and $\bm{\beta} \perp \bm{H}$ in (c), where the strength of the Ising SOI is fixed at $\beta = \Delta_{0}$.
	The critical field $H_c(T)$ is obtained from the data in (a) for each temperature. 
  } \label{fig1}
\end{figure*}

We solve the gap equation to obtain the transition temperature and draw the $H$-$T$ phase diagram.
The gap equation is expressed as
\begin{align} \label{eq:gap_equation}
  \Delta
  =
  - \frac{1}{4} g N_{0} T \sum_{\omega_{n}} \int d \xi \,
  \mathrm{Tr} \left[
    \check{\mathcal{F}} (\xi, \omega_{n}) (- i \hat{\sigma}_{y}) \hat{\rho}_{x}
  \right] ,
\end{align}
where  $N_{0}$ is the density of states at the Fermi level in the normal state
and $\mathrm{Tr}$ meaning the trace taken over spin and valley spaces 
extracts the spin-singlet even-valley symmetry correlation from the anomalous Green's function.
Namely, the attractive interaction $g > 0$ between two electrons works only at spin-singlet even-valley symmetry channel.
The transition temperature $T_c$ for $\bm{\beta} \perp \bm{H}$ is obtained 
by solving the linearized gap equation~\cite{frigeri:prl2004,xi:natphys2016,ilic:prl2017,mockli:prb2019,mockli:prb2020,roy:2dmate2025,ma:arXiv2025}
\begin{align} \label{eq:gapeq_at_Tc}
  0 = 
  \ln \frac{T}{T_{0}}
  +
  2 \pi T \sum_{\omega_{n} > 0} 
  \frac{1}{\omega_{n}} \left[
    1
    -
    \frac{\omega^{2}_{n} + \beta^{2}}{\omega^{2}_{n} + \beta^{2} + \mu_{\mathrm{B}}^2H^{2}} 
  \right],
\end{align}
with $T=T_c$, 
where $T_{0}$ is the transition temperature in the absence of a Zeeman field and the SOI.
In the limit of strong SOI $\beta \gg \mu_{\mathrm{B}} H$, 
it has been already known that the transition temperature approaches $T_{0}$.
The numerical results for the critical Zeeman field
$H_{c}$ are plotted as a function of temperature in Fig.~\ref{fig1} (a) 
for several choices of $\beta$.
The critical field $H_{c}$ and a temperature respectively 
are normalized to $\Delta_{0}$ and $T_{0}$, 
where $\Delta_{0}$ denotes the pair potential at $H = \beta = T = 0$.
At $\beta = 0$, the results for $T < 0.556 T_0$ correspond to the first order transition points 
to the superconducting phase~\cite{sarma:jpcs1963,maki:ptp1964}.
The arrow at the vertical axis indicates the Pauli limit $\mu_{\mathrm{B}} H_{p} = \Delta_{0} / \sqrt{2}$.
For $\beta \geq \Delta_{0}$, all the results are obtained from the second order transition points 
to the superconducting phase.
The critical fields increase with increasing $\beta$ 
in agreement with several previous studies~\cite{lu:science2015,saito:natphys2016,xi:natphys2016,barrera:natcom2018}.
This effect is referred to in recent contexts as the Ising protection of superconductivity. 
The results in Fig.~\ref{fig1} (a) show that the Ising protection becomes stronger 
at lower temperatures.
As shown in Appendix~\ref{sec:Hc_zero_temperature},
the critical fields diverge to infinity at zero temperature
as long as the phase transition to superconducting phase is continuous.

For a magnetic configuration of $\bm{\beta} \parallel \bm{H}$,
the gap equation~\eqref{eq:gap_equation} becomes
\begin{align}
  &\Delta
  =
  \frac{\Delta}{4} g N_{0} T \sum_{\omega_{n}} \int d \xi \,
  \mathrm{Tr} \left[ \check{Z}_{+}^{-1} + \check{Z}_{-}^{-1} \right] , \\
  &\check{Z}_{\pm} \nonumber \\
  &=
  \hat{\sigma}_0 \left\{ - (\xi \hat{\rho}_0 \pm \beta \hat{\rho}_{z})^{2} 
  + (\omega \pm i \mu_{\mathrm{B}} H)^{2} \hat{\rho}_0 + \Delta^{2} \hat{\rho}_0 \right\},
\end{align}
as shown in Appendix~\ref{sec:solution}.
By taking the trace over two valleys and shifting the integration variable 
as $\xi \pm \beta \to \xi$, the $\beta$-dependence is eliminated from the gap equation.
At $\bm{\beta}\times \bm{H} =0$, the SOI modifies the band dispersion only slightly.  
Therefore, the SOI does not affect the critical field for $\bm{\beta} \parallel \bm{H}$~\cite{roy:2dmate2025}.
The resulting $H_{c}$ is identical to that at $\beta = 0$ as plotted by a dashed 
line in Fig.~\ref{fig1}~(a). Thus, the magnetic configuration 
$\bm{\beta}\times \bm{H}$ is a source of the Ising protection.

The orbital effects due to a magnetic field explains the anisotropy of the 
Ising protection observed in 
experiments~\cite{lu:science2015,saito:natphys2016,xi:natphys2016,barrera:natcom2018}.
According to our analysis, 
the anisotropy of the Ising protection can be explained solely by the Zeeman effect.
In the next section, we explain mechanisms behind the anisotropy of the Ising protection and its low-temperature anomaly.

\section{Superfluid density} \label{sec:superfluid_density}
The transition temperature shown in Fig.~\ref{fig1}~(a) is obtained by solving the gap equation, 
where only spin-singlet pairing correlation contributes to the pair potential.
Here we discuss how three induced spin-triplet Cooper pairs
in Table~\ref{table1} stabilize or destabilize the superconducting states. 
To clarify the mechanism of the anisotropy and the low-temperature anomaly of Ising protection, 
we analyze the stability of the superconducting states 
by calculating the superfluid density~\cite{AGD} defined as
\begin{align}
  Q &= Q_{\mathcal{G}} + Q_{\mathcal{F}} , \\
  Q_{\mathcal{G}}
  &=
  \frac{1}{2} n T\sum_{\omega_{n}} \int d \xi \,
  \mathrm{Tr} \big[
    \check{\mathcal{G}} \check{\mathcal{G}} - \check{\mathcal{G}}^{\mathrm{N}} \check{\mathcal{G}}^{\mathrm{N}} 
    \big]_{(\bm{k}, \omega_{n})}
  , \\
  Q_{\mathcal{F}}
  &=
  \frac{1}{2} n T\sum_{\omega_{n}} \int d \xi \, \mathrm{Tr} \big[
    - \undertilde{\check{\mathcal{F}}} \check{\mathcal{F}}
  \big]_{(\bm{k}, \omega_{n})} ,
\end{align}
where $n$ is the electron density per spin and $\check{\mathcal{G}}^{\mathrm{N}}$ is the Green's function in the normal state.
As we find a relation $Q_{\mathcal{G}} = Q_{\mathcal{F}}$ as a result of the calculation, we discuss how the four pairing correlations in Table~\ref{table1} contribute to $Q_{\mathcal{F}}$.
Since $Q \propto \Delta^{2}$, 
the temperature dependence of $\Delta^{2}$ strongly governs the temperature dependence of $Q$.
Therefore, we evaluate the superfluid weight \textsl{along the phase transition line} in Fig.~\ref{fig1}~(a) 
which is defined as
\begin{align}
  q (T)
  =
  \frac{Q(H_{c} (T), T)}{n \Delta^{2}} \bigg|_{\Delta \to 0} .
\end{align}
Here $H_{c} (T)$ is determined from the data points on the phase boundary in Fig.~\ref{fig1}~(a).
From the superfluid weight,
we capture the contributions of the even- and odd-frequency pairing correlations
to the superfluid density just below the transition temperature.

Before going into details, we summarize the relations between the superfluid weights and the thermal
properties of superconductors. 
The thermodynamics of a superconductor near the transition temperature is well described by the Ginzburg-Landau free-energy
\begin{align} \label{eq:GL_free_energy}
  F_{\mathrm{SN}}
  =
  a \Delta^{2} + b \Delta^{4} + \mathrm{h.o.t.} .
\end{align}
The coefficient $a$ is proportional to the linearized gap equation 
shown in the right-hand-side of Eq.~\eqref{eq:gapeq_at_Tc}
which is 0 at $T = T_c$ and negative for $T < T_c$.  
The spin-singlet pairing correlation function changes its amplitude as a result of 
the appearance of three spin-triplet pairing correlations in Table 1. 
The suppression of $T_c$ represents the instability of the superconducting state quantitatively. 
The superfluid weight $q_{\mathcal{F}}$ tells the reasons and the quality of the instability.
The appearance of odd-frequency Cooper pairs destabilizes the superconducting states because 
they decrease the superfluid weight. 
It is often said that odd-frequency pairs exhibit a paramagnetic response, because the 
screening length of an external magnetic field obeys the relation $\lambda \propto q^{-1/2}$.
When $\lambda$ increases due to the appearance of odd-frequency pairs, the diamagnetism of the superconductor 
consequently weakens.
Actually, $q_{\mathcal{F}}$ is proportional to the coefficient $b$ with the same sign~\cite{sato:prb2024}.
Therefore, the condition $q_{\mathcal{F}} > 0$ ensures a second-order transition to the stable superconducting state.
Otherwise, the transition becomes first-order.

\subsection{Absence of a Zeeman field}\label{subsec:vzero}
In the absence of Zeeman fields, the SOI generates only the second term in Eq.~\eqref{eq:F}.
The gap equation in Eq.~\eqref{eq:gapeq_at_Tc} indicates $T_c=T_0$ 
irrespective of the amplitude of the SOI.
Here we discuss the roles of the pairing correlations proportional to $\hat{\xi}_{\bm{k}}$.
The superfluid density at $\bm{H}=0$ can be calculated exactly as
\begin{align}
 Q_{\mathcal{F}} &= Q_{s} + Q_{\beta} , \\
   Q_{s} &=
  2 \pi T \sum_{\omega_{n} > 0} \frac{n \Delta^2 (2\Omega^2 +\beta^2)}{2\Omega^3 (\Omega^2+\beta^2)}, \\
   Q_{\beta} &=
  2 \pi T \sum_{\omega_{n} > 0} \frac{n \Delta^2\, \beta^2 }{2\Omega^3 (\Omega^2+\beta^2)}, 
\end{align} 
with $\Omega=\sqrt{\omega_n^2+ \Delta^2 }$, where $Q_{s}$ and $Q_{\beta}$ are the superfluid density 
originate from the first and second terms in Eq.~\eqref{eq:F}, respectively. 
As a consequence, we find
\begin{align}
  Q_{s} + Q_{\beta} = Q_{\mathrm{BCS}} = 2 \pi T \sum_{\omega_{n} > 0} \frac{n \Delta^2}{\Omega^3},
\end{align}
where $Q_{\mathrm{BCS}}$ is the superfluid density in BCS theory.
Although $Q_s$ decreases with increasing $\beta$, the total 
superfluid density $Q_{\mathcal{F}}$ remains unchanged from $Q_{\mathrm{BCS}}$.
The pairing correlations proportional to ${\xi}_{\bm{k}}$ do not 
change the thermal properties of superconducting states.
Therefore, in what follows, 
we consider their combined superfluid density as $Q_0 = Q_{s} + Q_{\beta}$.

\subsection{Parallel configuration}
We next discuss the parallel configuration $\bm{\beta} \parallel \bm{H}$ in which  
the Ising protection is absent as shown in Fig.~\ref{fig1}~(a). 
The superfluid weight consists of two contributions,
\begin{align}
  q_{\mathcal{F}} &= q_{\mathrm{even}} + q_{\mathrm{odd}} , \quad q_{\mathrm{even}}=q_0,  \label{eq:q_comp1}\\
  q_{\mathrm{even}} 
  &=
  2 \pi T \sum_{\omega_{n} > 0}
  \frac{2 \omega^{4}_{n} - \omega^{2}_{n} \mu^{2}_{\mathrm{B}} H^{2}_{c} + \mu^{4}_{\mathrm{B}} H^{4}_{c}}
  {2 \omega_{n} (\omega^{2}_{n} + \mu^{2}_{\mathrm{B}} H^{2}_{c})^{3}} , \\
  q_{\mathrm{odd}}
  &=
  - 2 \pi T \sum_{\omega_{n} > 0}
  \frac{5 \omega^{2}_{n} \mu^{2}_{\mathrm{B}} H^{2}_{c} + \mu^{4}_{\mathrm{B}} H^{4}_{c}}
  {2 \omega_{n} (\omega^{2}_{n} + \mu^{2}_{\mathrm{B}} H^{2}_{c})^{3}} ,
\end{align}
where $q_{\mathrm{even}}$ and $q_{\mathrm{odd}}$ are the superfluid weight of even- and odd-frequency 
Cooper pairs, respectively. The superfluid weight of odd-frequency pairs is always negative.
The calculated results are plotted in Fig.~\ref{fig1}~(b).
The superfluid weight of even-frequency spin-singlet $s$-wave component $q_{\mathrm{even}}$ 
shows a nonmonotonic dependence on temperatures. 
For $T < 0.556 T_{0}$, the amplitude of the odd-frequency component becomes larger than that of
the even-frequency component, which leads to $q_{\mathcal{F}}<0$. 
As a result, the transition to the superconducting state 
becomes discontinuous~\cite{sato:prb2024}.

For the latter convenient, 
we supply the analytical results at low temperatures, where 
the Matsubara summation is replaced by the integration with
a low energy cut-off $\omega_{\mathrm{min}}= 2 \pi T$.
The superfluid weight is calculated to be
\begin{align}
  q_{\mathrm{even/odd}} 
  &\approx  +/-
  \frac{1}{2 \mu^{2}_{\mathrm{B}} H^{2}_{c}}
  \log\left(\frac{ \mu_{\mathrm{B}} H_c}
  {2 \pi T}\right)
  , \label{eq:qeo_parallel}
\end{align}
for $ T \ll \mu_{\mathrm{B}} H$.
Both the weights exhibit a logarithmic dependence on temperature.
It would be helpful to compare
the results in Eq.~\eqref{eq:qeo_parallel} with those in the superfluid weight in BCS theory
\begin{align}
  q_{\mathrm{BCS}} 
  &\approx
  \frac{1}{2 (2 \pi  T)^{2}}. \label{eq:qbcs}
\end{align}
The vertical axis in Fig.~\ref{fig1}~(b) and~(c) is normalized to $q_{\mathrm{BCS}}$ at $T=T_{0}$.
At low temperatures, $q_{\mathrm{BCS}}$ indicate the power-law dependence on temperature.
Thus, the expected relationship $q_{\mathrm{even}}  \ll q_{\mathrm{BCS}}$ 
also indicates the instability of the superconducting state.
A Zeeman field for $\bm{\beta}\parallel \bm{H}$ 
reduces the superfluid weight $q_{\mathrm{even}}$ drastically.

\subsection{Perpendicular configuration}
For $\bm{\beta} \perp \bm{H}$, 
the superfluid weight $q_{\mathcal{F}}$ consists of the three parts as
\begin{align}
  q_{\mathcal{F}} &= q_{\mathrm{even}} + q_{\mathrm{odd}} , \quad q_{\mathrm{even}}= q_{0} + q_{\perp} \label{eq:q_comp2} \\
  q_{0}
  &=
  2 \pi T \sum_{\omega_{n} > 0} \frac{1}{z} (\omega_n^2 + \beta^2)  
  \nonumber\\ &\times  
  \Big\{ 2 D_0^2 -\mu^{2}_{\mathrm{B}} H^{2}_{c}(4\omega_n^2 + D_0)  \Big\}  , \\
  q_{\mathrm{odd}}
  &=
  - 2 \pi T \sum_{\omega_{n} > 0} \frac{1}{z} \omega^{2}_{n} \mu^{2}_{\mathrm{B}} H^{2}_{c} 
  \Big( 4 \omega^{2}_{n} + D_0  \Big)
  , \label{eq:odd_perp}\\
  q_{\perp}
  &=
  2 \pi T \sum_{\omega_{n} > 0} \frac{1}{z} \beta^{2} \mu^{2}_{\mathrm{B}} H^{2}_{c}
  \Big( 4 \omega^{2}_{n} + D_0 \Big)
  , \\
  z &= 2 \omega^{3}_{n} D_0^3, \quad D_0=(\omega^{2}_{n} + \beta^{2} + \mu^{2}_{\mathrm{B}} H^{2}_{c}) .
\end{align}
The first term $q_{0}$ is derived from 
a spin-singlet pairing correlation linked to the pair potential
and the induced pairing correlation proportional to $\hat{\xi}_{\bm{k}}$ 
as discussed in Sec.~\ref{subsec:vzero}.
The component from the induced odd-frequency pairing correlation $q_{\mathrm{odd}}$ 
decreases the superfluid density~\cite{asano:prl2011,suzuki:prb2014,asano:prb2015,sato:prb2024}.
The term $q_{\perp}$ is the contributions of the induced even-frequency spin-triplet pairing correlations 
listed at the bottom row in Table~\ref{table1}.
We plot the superfluid weights as a function of temperature in Fig.~\ref{fig1} (c).
In contrast to the results in Fig.~\ref{fig1} (b), 
the amplitude of even-frequency components $q_{\mathrm{even}}=q_{0}+q_{\perp}$ is 
much larger than $q_{\mathrm{odd}}$, which indicates stable superconducting state.
The SOI relaxes the instability of the superconducting state due to a Zeeman field for $\bm{\beta} \perp \bm{H}$.
In particular at $T=0$, the equation $q_{\mathrm{odd}} / q_{\mathrm{even}} \approx 0$ 
leads to $\mu_{\mathrm{B}}H_c \gg \beta $, which explains the low-temperature anomaly in $H_c$.

The source of the Ising protection $q_{\mathrm{even}} \gg  q_{\mathrm{odd}}$ is 
explained well by the analytical 
expression of superfluid weights at low temperatures
\begin{align}
  q_{\mathrm{even}} 
  &\approx
 \frac{\mu^{2}_{\mathrm{B}} H^{2}_{c}}{2 (\beta^{2} + \mu^{2}_{\mathrm{B}} H^{2}_{c})^{2}}
 \ln \left( \frac{\sqrt{\beta^{2} + \mu^{2}_{\mathrm{B}} H^{2}_{c}}}{2\pi T} \right) \nonumber \\
 & \qquad + 
   \frac{1}{2}\left[ 1 - \frac{\mu^{2}_{\mathrm{B}} H^{2}_{c}}{\beta^{2} + \mu^{2}_{\mathrm{B}} H^{2}_{c}} 
   \right]\frac{1}{(2\pi T)^2} \label{eq:qeven_perp}
  , \\
  q_{\mathrm{odd}}
  &\approx
  - \frac{\mu^{2}_{\mathrm{B}} H^{2}_{c}}{2 (\beta^{2} + \mu^{2}_{\mathrm{B}} H^{2}_{c})^{2}}
  \ln \left( \frac{\sqrt{\beta^{2} + \mu^{2}_{\mathrm{B}} H^{2}_{c}}}{2 \pi T} \right) .\label{eq:qodd_perp}
\end{align}
The two superfluid weights exhibit qualitatively different behaviors from each other.
Equation~\eqref{eq:qeven_perp} shows that $q_{\mathrm{even}}$ recovers a power-law dependence on temperature as that in 
$q_{\mathrm{BCS}}$ as shown in Eq.~\eqref{eq:qbcs}.
On the other hand, $q_{\mathrm{odd}}$ remains a logarithmic dependence.
As shown in Eq.~\eqref{eq:F}, the odd-frequency pairing correlation is proportional to the Matsubara frequency 
$\omega_n$. As a result, $\omega_n^2$ appearing at the numerator of Eq.~\eqref{eq:odd_perp} 
changes the singularity of the integrand at small $\omega_n$ from $\omega_n^{-3}$ to $\omega_n^{-1}$. 
Equations~\eqref{eq:qeven_perp} and \eqref{eq:qodd_perp} mathematically express the physics behind 
the Ising protection of superconducting state.

\subsection{Intermediate configuration} \label{sec:anisotropy}
\begin{figure}[tttt]
  \centering
  \includegraphics[width = 0.7\columnwidth]{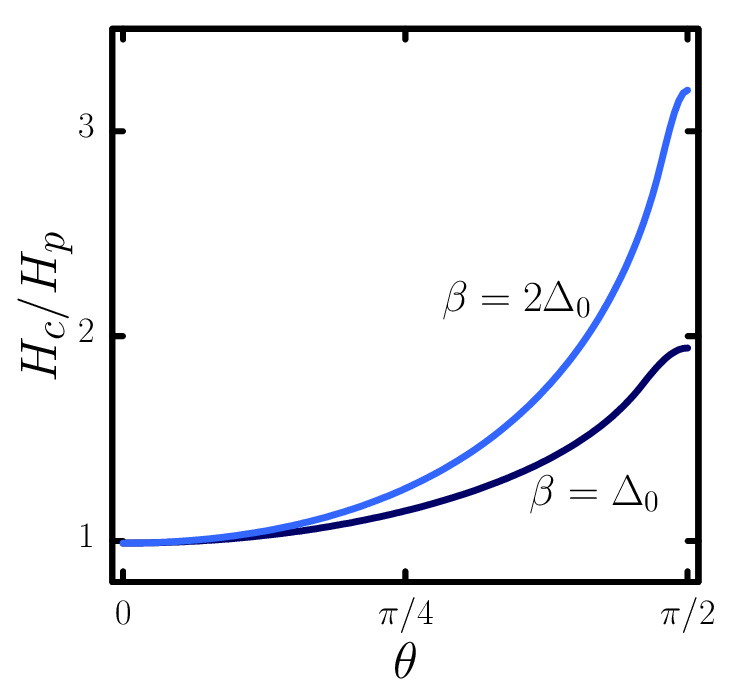}
  \caption{
    The critical Zeeman field $H_{c}$ at $T = 0.1 T_{0}$ is plotted as a function of the 
	  angle $\theta$ between $\bm{\beta}$ and $\bm{H}$ for two values of the 
	  Ising spin-orbit interactions $\beta$.
    The angles $\theta = 0$ and $\theta = \pi / 2$ correspond to the out-of-plane and in-plane magnetic fields, respectively.
  } \label{fig2}
\end{figure}

For intermediate configurations, we only show the numerical results of $H_c$ at 
$T = 0.1 T_{0}$ in Fig.~\ref{fig2}, where 
$\theta$ is the angle between $\bm{\beta}$ and $\bm{H}$. 
At $\theta = 0$ corresponding to the configuration $\bm{\beta} \parallel \bm{H}$, 
$H_c$ takes a value near the Pauli limit $H_{p}$ independent of $\bm{\beta}$. 
The critical field increases monotonically with increasing $\theta$.
At $\pi/2$ corresponding to $\bm{\beta} \perp \bm{H}$, $H_c$ becomes larger for larger $\beta$.

The shape of curves in Fig.~\ref{fig2} represents 
the degree of anisotropy solely from the Ising protection.
In experiments, however, $H_c$-$\theta$ curves are influenced by other factors.
At $\theta = 0$, $H_c$ in real TMDs is expected to be smaller than $H_{p}$ because the orbital
effect due to a magnetic field suppresses superconductivity.
The enhancement of $H_c$ around $\theta = \pi/2$ depends sensitively on temperature.
If the Rashba SOI coexists with the Ising SOI in real TMDs, the Rashba SOI drastically 
suppresses the enhancement of $H_c$ around $\theta = \pi/2$ 
as briefly discussed in Sec.~\ref{sec:discussion}.

\section{Discussion} \label{sec:discussion}
The enhancement of the critical field by impurities involving spin-orbit interaction 
has been established in previous studies~\cite{maki:ptp1964,klemm:prb1975}.
In these studies, impurity self-energy of an electron has 
the same spin-structure as that of Ising SOI 
because the motion of an electron is limited within a two-dimensional superconducting layer.
Thus, the impurity self-energy is considered 
to generate pairing correlation that belongs to the same symmetry class 
as the last term in Eq.~\eqref{eq:F}.
In fact, the gap equation for $T_{c}$ in the dirty limit 
takes a similar form to Eq.~\eqref{eq:gapeq_at_Tc}.

The Rashba SOI is another type of spin-orbit interaction.
The low-temperature anomaly of $H_{c}$ cannot be seen 
in $H$-$T$ phase diagram for a SC with the Rashba SOI~\cite{yuan:pnas2021}.
Unlike the Ising SOI, the Rashba SOI has two components: 
one is perpendicular to the Zeeman field and the other is parallel to the Zeeman field.
Although the perpendicular component stabilizes the superconducting state, 
the parallel component generates additional odd-frequency pairing correlations proportional to
\begin{align}
i \omega_n \, \mu_{\mathrm{B}} \bm{H} \cdot \bm{\beta} \, i \hat{\sigma}_y \, \hat{\rho}_y ,
\end{align}
according to Eq.~\eqref{eq:z_def}.
Thus, the perpendicular and parallel components of Rashba SOI 
exhibit opposite effects on $H_{c}$.
Both the Zeeman field and the Rashba SOI induce equal-spin Cooper pairs.
However, the stability of the superconducting state
is determined entirely by the frequency symmetry of the induced Cooper pairs,
regardless of their specific spin structures.
A previous numerical simulation~\cite{saito:natphys2016} has reported that 
the Rashba SOI drastically suppresses the low-temperature anomaly of $H_{c}$ 
due to the Ising SOI.
Their results demonstrate that the low-temperature anomaly disappears
when the amplitude of Rashba SOI reaches 6\% of the Ising SOI.

\section{Conclusion} \label{sec:conclusion}
We have theoretically investigated the mechanism of the Ising protection of superconducting 
state in monolayer transition-metal dichalcogenide superconductors under a Zeeman field parallel to the layer.
The following conclusions can be obtained from the analytical expression 
of the anomalous Green's function and that of the superfluid density. 
In a spin-singlet superconductor, a Zeeman field $\bm{H}$ always makes the superconducting state unstable because it 
reduces the superfluid density by breaking the even-frequency Cooper pairs and 
by generating odd-frequency Cooper pairs. 
Generally speaking, even-frequency (odd-frequency) Cooper pairs increase (decrease) superfluid density. 
The Ising spin-orbit interaction $\bm{\beta}\perp \bm{H}$ drastically increases 
the superfluid density and
relaxes the instability due to a Zeeman field. Such effect is more remarkable in lower temperatures.
However, the Ising spin-orbit interaction $\bm{\beta}\parallel \bm{H}$ does not change 
the superfluid density at all. 
These characteristic behaviors of the superfluid densities explains the 
anisotropy of the Ising protection and its low-temperature anomaly. 
Such an anisotropy is explained also by the presence of 
induced even-frequency Cooper pairs whose correlation function is proportional to $\bm{\beta} \times \bm{H}$.

\section*{Acknowledgments}
T. S. was supported by JST SPRING, Grant Number JPMJSP2119.
Y. A. is supported by a Grant-in-Aid for Scientific Research (JSPS KAKENHI Grant No. JP26K0692).

\appendix

\begin{widetext}
\section{SOLUTION}\label{sec:solution}
In this appendix, we show the full BdG Hamiltonian and the Green's functions.
The BdG Hamiltonian reads
\begin{align}
H_{\mathrm{BdG}}(\bm{k})  \label{eq:full BdG Hamiltonian}
&=
\begin{bmatrix}
  \xi_{-} \hat{\sigma}_0 + \bm{\beta} \cdot \hat{\bm{\sigma}} + \bm{V} \cdot \hat{\bm{\sigma}} & 0 & 
  0 & \Delta i \hat{\sigma}_{y} \\
  0 & \xi_{+} \hat{\sigma}_0 - \bm{\beta} \cdot \hat{\bm{\sigma}} + \bm{V} \cdot \hat{\bm{\sigma}} &
  \Delta i \hat{\sigma}_{y} & 0 \\
  0 & - \Delta i \hat{\sigma}_{y} &
  - \xi_{+} \hat{\sigma}_0 - \bm{\beta} \cdot \hat{\bm{\sigma}}^{*} - \bm{V} \cdot \hat{\bm{\sigma}}^{*} & 0 \\
  - \Delta i \hat{\sigma}_{y} & 0 &
  0 & - \xi_{-} \hat{\sigma}_0 + \bm{\beta} \cdot \hat{\bm{\sigma}}^{*} - \bm{V} \cdot \hat{\bm{\sigma}}^{*}
\end{bmatrix} , \\
\xi_{\pm} 
&= 
\frac{1}{2 m}(\bm{k} \pm \bm{K})^2 - \mu, \quad
\bm{V} = \mu_{\mathrm{B}} \bm{H}.
\end{align}
The solution of the Gor'kov equation is calculated as
\begin{align}
  &\check{\mathcal{G}} (\bm{k}, \omega_{n}) \nonumber \\
  &=
  \check{Z}^{-1} \left[
    \left\{
      \left( i \omega_{n} \hat{\rho}_{0} - \hat{\xi} \right) \hat{z}_{\mathrm{N}}
      -
      \left( i \omega_{n} \hat{\rho}_{0} + \hat{\xi} \right) \Delta^{2}
    \right\} \hat{\sigma}_{0}
    +
    \left( \hat{z}_{\mathrm{N}} + \Delta^{2} \hat{\rho}_{0} \right) \bm{V} \cdot \hat{\bm{\sigma}}
    +
    \left( \hat{z}_{\mathrm{N}} - \Delta^{2} \hat{\rho}_{0} \right) \bm{\beta} \cdot \hat{\bm{\sigma}} \hat{\rho}_{z}
  \right] , \\
  &\check{\mathcal{F}} (\bm{k}, \omega_{n}) \nonumber \\
  &=
  \check{Z}^{-1} \left[
    - \left\{ 
      \hat{\xi}^{2} + (\omega^{2}_{n} + \Delta^{2} + \bm{\beta}^{2} - \bm{V}^{2}) \hat{\rho}_{0}
    \right\} \hat{\sigma}_{0}
    +
    2 i \omega_{n} \bm{V} \cdot \hat{\bm{\sigma}} \hat{\rho}_{0}
    +
    2 \hat{\xi} \, \bm{\beta} \cdot \hat{\bm{\sigma}} \hat{\rho}_{z}
    - 
    2 i \left( \bm{V} \times \bm{\beta} \right) \cdot \hat{\bm{\sigma}} \hat{\rho}_{z}
  \right]
  \Delta \left( i \hat{\sigma}_{y} \right) \hat{\rho}_{x} , \\
  &\check{Z}
  =
  \hat{\sigma}_0 \Big[
    \left\{
      \hat{\xi}^{2} + (\omega^{2}_{n} + \Delta^{2} + \bm{\beta}^{2} - \bm{V}^{2}) \hat{\rho}_{0}
    \right\}^{2}
    -
    4 \left(
      i \omega_{n} \bm{V} \hat{\rho}_{0} + \hat{\xi} \bm{\beta} \hat{\rho}_{z}
    \right)^{2}
    +
    4 \left( \bm{V} \times \bm{\beta} \right)^{2} \hat{\rho}_{0}
  \Big] , \\
  &\hat{z}_{\mathrm{N}}
  =
  \left( i \omega_{n} \hat{\rho}_0 + \hat{\xi} \right)^{2}
  -
  \left( \bm{V} \hat{\rho}_0 - \bm{\beta} \hat{\rho}_{z} \right)^{2} , \\
  &\hat{\xi}
  =
  \begin{bmatrix} \xi_{-} & 0 \\ 0 & \xi_{+} \end{bmatrix} .
\end{align}

The anomalous Green's function for $\bm{\beta} \parallel \bm{H} \parallel \bm{z} $ 
near the transition temperature is represented as
\begin{align}
  \check{\mathcal{F}} (\bm{k}, \omega_{n})
  =&-\frac{1}{2} \left[
  \left\{(\hat{\xi}-\beta \hat{\rho}_z)^2 +(\omega_n-iV)^2 \hat{\rho}_0 \right\}^{-1} \hat{\sigma}_0
  + \left\{(\hat{\xi}+\beta \hat{\rho}_z)^2 +(\omega_n+iV)^2 \hat{\rho}_0 \right\}^{-1} \hat{\sigma}_0 \right.\nonumber\\
 &+ \left.\left\{(\hat{\xi}-\beta \hat{\rho}_z)^2 +(\omega_n-iV)^2 \hat{\rho}_0 \right\}^{-1} \hat{\sigma}_z
  - \left\{(\hat{\xi}+\beta \hat{\rho}_z)^2 +(\omega_n+iV)^2 \hat{\rho}_0 \right\}^{-1} \hat{\sigma}_z
  \right]\Delta \left( i \hat{\sigma}_{y} \right) \hat{\rho}_{x} .
\end{align}
The gap equation is composed of the summation of the anomalous Green's function over wavenumbers.
\begin{align}
  &\frac{1}{V_{\mathrm{vol}}} \sum_{\bm{k}} 
  \check{\mathcal{F}} (\bm{k}, \omega_{n}) \nonumber \\
  &=-\frac{1}{2} 
  \frac{1}{V_{\mathrm{vol}}} \sum_{\bm{k}} 
  \left[
  \left\{({\xi}_{\bm{k}} \hat{\rho}_0 -\beta \hat{\rho}_z)^2 +(\omega_n-iV)^2 \hat{\rho}_0 \right\}^{-1} \hat{\sigma}_0 
  + \left\{({\xi}_{\bm{k}} \hat{\rho}_0+\beta \hat{\rho}_z)^2 +(\omega_n+iV)^2 \hat{\rho}_0 \right\}^{-1} \hat{\sigma}_0 \right.\nonumber\\
 &\qquad 
 + \left.\left\{({\xi}_{\bm{k}} \hat{\rho}_0-\beta \hat{\rho}_z)^2 +(\omega_n-iV)^2 \hat{\rho}_0 \right\}^{-1} \hat{\sigma}_z
  - \left\{({\xi}_{\bm{k}} \hat{\rho}_0 +\beta \hat{\rho}_z)^2 +(\omega_n+iV)^2 \hat{\rho}_0 \right\}^{-1} \hat{\sigma}_z
  \right]\Delta \left( i \hat{\sigma}_{y} \right) \hat{\rho}_{x},\label{eq:aa1}\\
  &=-\frac{1}{2} 
  N_0 \int d\xi 
  \left[
  \frac{1}{ \xi^2 +(\omega_n-iV)^2 } \hat{\sigma}_0 + \frac{1}{ \xi^2 +(\omega_n+iV)^2 } \hat{\sigma}_0
 \right.\nonumber\\
 &\qquad  +\left.\frac{1}{ \xi^2 +(\omega_n-iV)^2 }\hat{\sigma}_z 
 - \frac{1}{ \xi^2 +(\omega_n+iV)^2 } \hat{\sigma}_z
  \right]\Delta \left( i \hat{\sigma}_{y} \right) \hat{\rho}_{x},\label{eq:aa2}\\
&\xi_{\bm{k}} = \frac{\bm{k}^2}{2m} - \mu.
\end{align}
We shift 
the wavenumber as $\bm{k}\to \bm{k} \mp \bm{K}$ for $\xi_\pm$ to reach Eq.~\eqref{eq:aa1}
and change $\xi \pm \beta \to \xi$ to reach Eq.~\eqref{eq:aa2}.
Equation~\eqref{eq:aa2} has exactly the same expression as that for the Green's function 
in uniform Zeeman field. 
Thus, $T_c$ for $\bm{\beta} \parallel \bm{H}$ is equal to that with $\bm{\beta}=0$.
The first two terms in Eq.~\eqref{eq:aa2} represent even-frequency spin-singlet $s$-wave 
even-valley Cooper pairs. 
The last two terms are the correlation of odd-frequency spin-triplet $s$-wave even-valley
Cooper pairs. 
The two components contribute to the superfluid density in the opposite way
as shown in the text.

The anomalous Green's function for $\bm{\beta} \parallel  \bm{z} $ and $\bm{H} \parallel \bm{y}$
near the transition temperature is obtained
\begin{align}
  \frac{1}{V_{\mathrm{vol}}} \sum_{\bm{k}} 
  \check{\mathcal{F}} (\bm{k}, \omega_{n})
  =&- 
  \frac{1}{V_{\mathrm{vol}}} \sum_{\bm{k}} 
  \frac{\left[
  (\xi^2_{\bm{k}}+\omega^2+\beta^2-V^2) \hat{\sigma}_0 \hat{\rho}_0 + 2i \omega_n V \hat{\sigma}_y \hat{\rho}_0
  + 2 \xi_{\bm{k}} \beta \hat{\sigma}_z \hat{\rho}_z - 2 i V \beta \hat{\sigma}_x \hat{\rho}_z
  \right]}
  {\xi^4_{\bm{k}} +2 \xi^2_{\bm{k}} (\omega_n^2-\beta^2 - V^2 )+ (\omega_n^2+\beta^2 + V^2 )^2}
  \Delta \left( i \hat{\sigma}_{y} \right) \hat{\rho}_{x}.\label{eq:aa3}
\end{align}
Here we have already shifted the wavenumber $\bm{k} \to \bm{k}\mp \bm{K}$ for $\xi_\pm$.

\section{CRITICAL MAGNETIC FIELD AT ZERO TEMPERATURE} \label{sec:Hc_zero_temperature}
Here, we show that the critical Zeeman field diverges 
in the zero-temperature limit from the linearized gap equation.
In the low-temperature limit,
by using the relation for the summation of the Matsubara frequency 
\begin{align}
  2 \pi T \sum_{\omega_n > 0} \to 
  \int_{\omega_{\mathrm{min}} = 2 \pi T}^{\infty} d \omega 
\end{align}
the linearized gap equation is calculated to be
\begin{align}
  0 
  &= 
  \ln \frac{T}{T_0}  
  +
  \frac{\mu_{\mathrm{B}}^2 H^2}{\beta^2 + \mu_{\mathrm{B}}^2 H^2}
  \int^{\infty}_{\omega_{\mathrm{min}}} d \omega \,
  \left\{
    \frac{1}{\omega} 
    -
    \frac{\omega}{\omega^2 + \beta^2 + \mu_{\mathrm{B}}^2 H^2}
  \right\}, \nonumber \\
  &=
  \ln \frac{T}{T_0}
  +
  \frac{\mu_{\mathrm{B}}^2 H^2}{\beta^2 + \mu_{\mathrm{B}}^2 H^2}
  \left[
    \ln \omega
    - 
    \ln \sqrt{\omega^2 + \beta^2 + \mu_{\mathrm{B}}^2 H^2}
  \right]_{\omega_{\mathrm{min}}}^{\infty} , \\
  &=
  \ln \frac{T}{T_0}
  +
  \frac{\mu_{\mathrm{B}}^2 H^2}{\beta^2 + \mu_{\mathrm{B}}^2 H^2}
  \ln \left( \frac{{\beta^2 + \mu_{\mathrm{B}}^2 H^2}}{\omega_{\mathrm{min}}} \right) ,
\end{align}
where we have neglected $\omega_{\mathrm{min}}^2$
by assuming $\omega_{\mathrm{min}} = 2 \pi T \ll \beta, \, \mu_{\mathrm{B}} H$ 
in the low-temperature regime.
By replacing $\omega_{\mathrm{min}}$ with $2 \pi T$ and collecting the $\ln(2\pi T)$ terms, 
this equation can be rewritten as
\begin{align}
  \ln (2 \pi T)
  =
  \ln \left[
    2 \pi T_0 \left( 
      \frac{2 \pi T_0}{\sqrt{\beta^2 + \mu_{\mathrm{B}}^2 H^2}}
    \right)^{\frac{\mu_{\mathrm{B}}^2 H^2}{\beta^2}}
  \right] .
\end{align}
Taking the exponential of both sides yields the expression for temperature $T$
\begin{align}
  T \label{eq:Hc_zero temperature}
  =
  T_0 \exp \left[
    - \frac{\mu_{\mathrm{B}}^2 H^2}{\beta^2} 
    \ln \frac{\sqrt{\beta^2 + \mu_{\mathrm{B}}^2 H^2}}{2 \pi T_0}
  \right] .
\end{align}
This expression gives us the behavior of the critical magnetic field at zero temperature.
At zero temperature $T=0$,
the right-hand side of Eq.~\eqref{eq:Hc_zero temperature} must vanish.
Since $T_0$ is a non-zero positive constant, 
this condition requires the exponential factor to approach zero.
Only in the infinite-field limit ($H \to \infty$),
the exponent approaches negative infinity and the right-hand side becomes zero.
Therefore, to satisfy the linearized gap equation at $T=0$,
the critical magnetic field must diverge to infinity. 
This conclusion holds true as long as 
the phase transition to the superconducting phase is continuous, 
which means that the analysis based on the linearized gap equation 
remains valid within this regime.

\end{widetext}

\bibliography{list_ref}

\end{document}